# Metal-organic chemical vapor deposition of MgSiN$_2$ thin films

Vijay Gopal Thirupakuzi Vangipuram[1], Chenxi Hu[2], Abdul Mukit Majumder[1], Christopher Chae[3], Kaitian Zhang[1], Jinwoo Hwang[3], Kathleen Kash[2], Hongping Zhao[1,3,*]

[1] Department of Electrical and Computer Engineering, The Ohio State University, Columbus OH 43210

[2] Department of Physics, Case Western Reserve University, Cleveland OH 44106

[3] Department of Materials Science and Engineering, The Ohio State University, Columbus OH 43210

*Corresponding author email: zhao.2592@osu.edu

## Abstract

Orthorhombic-structured II-IV nitrides provide a promising opportunity to expand the material platform while maintaining compatibility with the wurtzite crystal structure of the traditional III-nitride material system. Among them, MgSiN$_2$ stands out due to its close compatibility with GaN and AlN and its theoretically predicted ultrawide direct band gap of 6.28 eV. In this work, the growth of MgSiN$_2$ thin films on GaN-on-sapphire and c-plane sapphire substrates was investigated using metal-organic chemical vapor deposition (MOCVD). MOCVD growth conditions were correlated with film quality and crystallinity for samples grown on GaN-on-sapphire substrates. The effects of Mg:Si precursor molar flow rate ratios and growth pressure at two different temperatures, 745°C and 850°C, were studied comprehensively. High-resolution scanning transmission electron microscopy (STEM) imaging confirmed the formation of high-quality, single-crystal MgSiN$_2$ films. Optical band gap extraction from transmittance measurements yielded direct band gap values ranging from 6.13 eV to 6.27 eV for samples grown under various conditions, confirming the realization of an ultrawide-band gap, III-nitride-compatible, II-IV-nitride material.

# I. Introduction

III-nitrides such as GaN, AlN and InN and their alloys have revolutionized the development of high-performance devices in multiple fields and application spaces, including high electron-mobility transistors (HEMTs), light-emitting diodes (LEDs) and laser diode technologies over the past few decades. Alloying has enabled tuning of the band gap and material properties over a wide range. However, there are still a number of challenges that limit current capabilities. For example, alloyed structures with high compositional variations are limited in film thickness due to large strain build-up if the lattice mismatch is large.[1] This limits heterostructure and heterojunction designs and thereby inhibits the development of devices with higher performance. II-IV-nitrides can serve as a crucial link in harnessing material properties that are otherwise not possible within traditional III-nitrides due to this issue. Utilizing elements from group II (e.g. Mg, Zn) and elements from group IV (e.g. Si, Ge, Sn) to occupy the group III lattices in an ordered cation sublattice that preserves the octet rule yields II-IV-nitrides with an orthorhombic crystal structure based on the wurtzite structure.[2] In addition to providing a potential pathway to overcoming current material limitations facing the III-nitrides, the II-IV-nitrides could provide new opportunities to achieve and utilize more unique material properties that have so far not been observed in traditional III-nitride materials, such as ferroelectricity and ferromagnetism.

Realizing single-crystal thin film growth of the II-IV-nitrides is a first step towards experimentally characterizing and fundamentally understanding the material properties of these materials. Development of metal-organic chemical vapor deposition (MOCVD) for them is particularly beneficial since it is the primary epitaxial growth method currently used in mass-manufacturing of III-nitride-based materials and devices.

Previous MOCVD growth studies have shown promising results. Thin single-crystal films of the II-IV-nitride compounds ZnGeN$_2$ and ZnSnN$_2$[3–5] and the alloys ZnGeGa$_2$N$_4$ and ZnSn(Ga)N$_2$[4–6] have been grown by MOCVD.

MgSiN$_2$ has been predicted to be an indirect band gap semiconductor with an indirect gap of 5.84 eV, and a direct gap of 6.28 eV.[2] The calculated *a* and *b* lattice constants of MgSiN$_2$ are mismatched to those of GaN by 1.41% and -4.1%, respectively, when transformed into the wurtzite crystal structure.[2] MgSiN$_2$ thus can be considered as a promising candidate to form heterojunctions with GaN. Theoretical predictions of alloyed compositions of MgSiN$_2$-GaN yield a direct band gap of 4.83 eV (1:1 of MgSiN$_2$/GaN), corresponding to a deep-UV wavelength of ~257 nm, and an indirect band gap less than 0.1 eV lower.[7] In addition, calculations also show MgSiN$_2$ should be ferroelectric, in contrast to the III-nitrides, with a predicted theoretical breakdown field higher than that of AlN, and a coercive field lower than the theoretical predictions for AlN. [8,9]

To date, MgSiN$_2$ has only been experimentally realized in powdered form.[10–13] It was shown to be stable and oxidation-resistant in ambient conditions up to 830°C.[10] Its reported chemical resistance to KOH solutions and mild etching in H$_3$PO$_4$ and HF solutions suggest good stability and etch selectivity. These are promising qualities when considering opportunities for future microfabrication and device processing of this material.[14]

This work reports on single-crystal MgSiN$_2$ thin films grown by MOCVD. The effects of the ratio of the Mg:Si precursor molar flow rate, chamber pressure, and growth temperature on composition, morphology and x-ray diffraction were investigated. Optical transmittance measurements were performed to extract the optical band gap.

## II. Experimental Details

Growth of the MgSiN$_2$ thin films was carried out in a commercial, vertical rotating-disk MOCVD reactor with N$_2$ used as the carrier gas. Silane was used as the Si precursor while bis(methylcyclopentadienyl)magnesium [(MeCp)$_2$Mg] was used as the metal-organic (MO) precursor for Mg. (MeCp)$_2$Mg was chosen because of its higher vapor pressure compared to the more commonly used Cp$_2$Mg. For all films reported in this work, the MO bath temperature was set to 40 °C and an estimated molar flow rate of 9.7 µmol/min. The Mg:Si molar flow rate ratio was tuned by varying the silane flow rate. The films were grown directly on Ga-polar, epitaxy-ready GaN-on-sapphire templates on a SiC-coated graphite susceptor. All samples in this work were grown for a duration of 1 hr. Table 1 lists the samples and their corresponding growth conditions. Samples A-D were grown at a temperature of 745°C, with variations in the Mg:Si precursor molar flow rate ratio. Samples E-H were grown at a temperature of 850°C, with variations in the Mg:Si precursor molar flow rate ratio. To investigate the effects of variations in growth pressure, Samples I-M and Samples N-R were grown at 745°C and 850°C, respectively, with growth pressures varying between 450 Torr and 550 Torr.

Scanning electron microscopy (SEM) was performed using a Thermo Scientific Apreo with an acceleration voltage of 30 kV and a beam current of 0.80 nA. Energy dispersive X-ray spectroscopy (SEM-EDX) was carried out on the same tool, with an acceleration voltage of 5 kV, to measure elemental compositions and Mg:Si composition ratios. To characterize the surface morphology, atomic force microscopy (AFM) was performed using a Bruker AXS Dimension Icon system. Scan areas of 5µm x 5µm, roughly ½" from the center of the 2" diameter wafers, were examined. X-ray diffraction (XRD) ω-2θ scans were performed on a Bruker D8 Discover system utilizing the Cu-K$_\alpha$ wavelength of 1.5416 Å. Cross-sectional lamellas for scanning transmission electron microscopy (STEM) were prepared using a Thermo Scientific Helios Nanolab 650 DualBeam focus ion beam (FIB). STEM was performed on a Thermo Scientific Themis Z S/TEM with an acceleration voltage of 300 kV. Optical transmission measurements from 1.0 eV to 6.7 eV photon energies were done with an Agilent Cary 5000 UV-VIS-NIR spectrophotometer.

## III. Results and Discussions

All samples discussed in Sections III.A to III.C were grown on epi-ready GaN-on-sapphire templates while samples discussed in Section III.D were grown on double-side polished, c-plane sapphire substrates. Sections III.A and III.B discuss the effects of variations in II/IV molar flow rate ratio and growth pressure on $MgSiN_2$ film properties. STEM analysis is included in Section III.C. Optical transmittance measurements and optical band gap extraction are discussed in Section III.D.

### III.A. Effects of variations in Mg:Si precursor molar flow rate ratio

The effects of varying the Mg:Si precursor molar flow rate ratio (FRR) were investigated at two different temperatures. Samples A-D were grown at 745°C with Mg:Si molar flow rate ratio varying from 3.22 to 3.46. Samples E-H were grown at 850°C with Mg:Si molar flow rate ratio varied between 8.20 and 9.00. Figure 1 shows the XRD ω-2θ scan, AFM imaging, SEM micrographs and the measured Mg:Si SEM-EDX ratios for the films grown at 745°C at different conditions. From XRD peak comparisons, it is evident that the strong substrate peak is present in all cases, but the (002) $MgSiN_2$ peak is not present for the conditions with the lowest (3.22) and highest (3.46) Mg:Si molar flow rate ratios, in which the films likely deviate from the ideal stoichiometry. The uncertainties in the EDX measurements of stoichiometry are approximately 9%. However, the XRD $MgSiN_2$ peak intensity varies from absent to prominent for a less than 3% change in the ratio of the Mg:Si precursor flow rates. Absent a more sensitive measurement of the stoichiometry, we suggest that the XRD measurement of the MgSiN$_2$ (002) peak is a more reliable indication of crystalline quality.

SEM imaging shows the presence of geometric surface features that seem to align with the hexagonal crystallographic axes of the substrate for Sample B and Sample C, indicating preferential crystalline growth has occurred for these conditions. AFM surface morphology of Sample C shows the same geometric surface features observed via SEM imaging. The surface is roughest for the case where the Mg:Si molar flow rate ratio is highest (Mg:Si molar flow rate ratio = 3.46) among this series, with an extracted RMS surface roughness of 11.8 ± 0.6 nm.

Samples E-H were grown with approximately 3.7% step increases in the Mg:Si molar flow rate ratio at the higher growth temperature of 850°C. Figure 2 shows the AFM surface morphology, SEM micrographs and XRD scan results for these samples. Here, unlike for the previous series, the (002) MgSiN$_2$ XRD peaks were observed for all four samples, with the highest intensity peaks observed for the samples grown with Mg:Si molar flow rate ratios of 8.50 and 8.70. While the presence of the XRD peaks for samples grown for this wider range of Mg:Si molar flow rate ratios indicates a larger growth window at this higher growth temperature, the AFM RMS surface roughness values are consistently higher than for those grown at 745°C. The lowest RMS roughness for this series was obtained for the lowest Mg:Si molar flow rate ratio (Sample E), although the XRD film peak intensity is weakest for this sample. Increased surface roughness and apparent island-like surface features are observed for the conditions where the XRD peak intensity was highest (Sample F and Sample G). These results imply that while better crystallinity is achieved under conditions corresponding to Sample F and Sample G, the slightly off-stoichiometry variations of Sample E and Sample H implied by the lower XRD peak intensities in fact yield smoother surfaces while not providing as high crystallinity. Compositional, stoichiometric variation is likely a contributing factor toward the presence of a secondary peak at about 36.5° in the XRD scan for Sample H.

No clear trends in the variation of Mg:Si compositional ratios from SEM-EDX analysis are observed across the series of samples investigated at both temperature conditions, despite changes in the Mg:Si molar flow rate ratio (Fig. 3 (a)). It is clear that, higher Mg:Si molar flow rate ratios are required to obtain close to stoichiometric compositions at higher growth temperatures. However, for both temperature conditions investigated, at the highest Mg:Si molar flow rate ratio (Sample D and Sample H), the extracted Mg:Si compositional ratios from SEM-EDX are higher than all other corresponding conditions, showing a clear increase in incorporated Mg under these conditions as compared to the conditions with lower Mg:Si molar flow rate ratios.

Figure 3(b) plots the ratio of the $MgSiN_2$ (002) XRD peak intensity to the GaN template (002) XRD peak intensity, obtained from ω-2θ scans, versus the corresponding Mg:Si molar flow rate ratio. From this graph, a growth window associated with the Mg:Si molar flow rate ratio is established at the two different temperature conditions investigated, with the highest XRD peak intensity ratios obtained being attributed to films closest to stoichiometry and highest crystallinity (and thus the highest XRD peak intensities). The graph also shows that a wider growth window is available with increases in temperature and changes in the Mg:Si molar flow rate ratio. At a growth temperature of 745°C, a change in the Mg:Si molar flow rate from 3.30 (Sample B) to 3.22 (Sample A), a change of approximately 2.4%, results in a significantly lower XRD peak intensity ratio, by a factor of almost 40. At a growth temperature of 850°C, a change in the Mg:Si molar flow rate ratio from 8.70 (Sample G) to 8.20 (Sample E), a change of approximately 6%, results in a change of a factor of approximately 13 in the XRD peak intensity ratio. Wider growth windows are advantageous in providing more robust film growth capabilities with opportunities for even further fine-tuning of the growth conditions as films are developed towards electronic and optoelectronic device applications.

### III.B. Effects of Variations in Pressure

Samples I-M were grown at a growth temperature of 745°C with a constant Mg:Si molar flow rate ratio of 3.30. The chamber pressure was varied between 450 Torr and 550 Torr in steps of 25 Torr. Under these conditions, the smoothest RMS surface roughness of $3.7 \pm 0.2$ nm was obtained for the film grown at 525 Torr (Sample L). Sample L also shows high XRD peak intensity for the (002) $MgSiN_2$ peak. Sample K, grown at 500 Torr, shows a prominent $MgSiN_2$ film XRD peak as well, although two additional shoulder peaks are also present, suggesting the possibility of stoichiometric variation or alternative crystal plane orientations. The SEM-EDX measured Mg:Si ratio is lower for Sample K compared to Sample L. At the highest and lowest pressure conditions investigated (Samples I and M, respectively), no obvious XRD peaks corresponding to $MgSiN_2$ are observed, even though the measured SEM-EDX Mg:Si stoichiometric ratios are close to ideal. This result suggests that, with appropriate changes in pressure, given a constant Mg:Si molar flow rate ratio, it is possible to transition from single-crystal oriented film of $MgSiN_2$ to non-crystalline structures, or mosaic films; that is, films with different crystallographic orientations present. For the samples that do not have an observable XRD peak (Samples I and M), the RMS surface roughness is larger than for the sample with a single, sharp XRD film peak (Sample L). With increases in pressure the gas phase reaction rates between precursor molecules are expected to increase. At the same time, higher pressures also influence the adatom mobilities on the growth surface. Reduced surface diffusion is expected at higher pressure due to increased molecular collisions within the gas phase, resulting in arriving adatoms possessing lower kinetic energies. The reduced surface diffusion of the arriving species results in faster adatom incorporation.[15-17]

The competition for incorporation between the arriving Mg and Si adatoms at the surface can strongly influence the crystal quality and stoichiometry of the film. Therefore, with the change in pressure, the films could have minimal or no perceived obvious trend with measured Mg:Si compositional ratio due to the complex interactions involved.

Samples N-R were grown at a growth temperature of 850°C with a constant Mg:Si molar flow rate ratio of 8.70, with the growth pressure again varied from 450 Torr to 550 Torr in 25 Torr steps. Unlike the films grown at the lower temperature of 745 °C, all films grown at 850 °C show (002) MgSiN$_2$ XRD peaks. These results imply a wider growth window at the higher growth temperature. Sample Q, grown at 525 Torr, shows the highest film peak intensity from ω-2θ scans when normalized to the substrate peak intensity. This sample also shows the presence of a secondary shoulder peak. The RMS surface roughnesses of all the films grown at 850°C are higher than those grown at 745°C. The samples grown at the highest and lowest pressures (Samples R and N, respectively) show a significant reduction in RMS surface roughness compared to the samples grown at intermediate pressures. No obvious correlations are observed between the measured SEM-EDX Mg:Si ratios and growth pressures. Figure 6(a) shows the (002) XRD peak intensities of the MgSiN$_2$ films, normalized to the GaN (002) peak intensities, plotted against the chamber pressure for Samples I-R. The highest XRD peak intensities were obtained for films grown at 745°C. There is a clear drop-off of the XRD peak intensity ratio at the lowest and highest pressures for these samples. The samples grown at 850 °C show much less variation in XRD peak intensities with pressure, suggesting that even wider growth windows are possible at higher temperatures. Figure 6(b) plots the RMS surface roughness. The films grown at 745 °C show the lowest RMS surface roughness for the samples with the highest XRD peak intensity ratios. In contrast, the films grown at 850 °C that have larger XRD peak intensity ratios are rougher.

### III.C. Scanning Transmission Electron Microscopy

High-angle annular dark-field (HAADF) STEM imaging on Sample B, grown at 745°C, is shown in Fig. 7(a). The cross section shows crystalline ordering within the MgSiN$_2$ film. While the crystalline nature is not apparent across the complete STEM image, it is currently unclear whether this is a result of surface oxidation on the prepared STEM sample, or a condition associated with oxidation or non-uniformity in the crystalline structure formed during the growth process. As seen in Fig. 7(b) from a large-area cross-sectional view, the resultant MgSiN$_2$ film grown on GaN is continuous across the complete substrate surface, with an average measured film thickness of 129 ± 7 nm. A STEM-EDX measurement across the MgSiN$_2$ cross-sectional area is shown in Fig. 7(c). The composition is fairly constant with depth and no obvious clustering of Mg, Si or N regions. EDX color mapping of the elements is shown in Fig. 7(d)[(i)-(vi)]. The GaN/MgSiN$_2$ interface is sharp, with no evidence of significant diffusion of any constituent elements either from the film into the substrate, or from the substrate into the film. The oxygen concentration is evident in regions in the MgSiN$_2$ near the film/substrate interface. These regions show higher Mg and lower Si concentrations, possibly a result of oxidation during initial film growth, during STEM sample preparation, or due to exposure to air.

The HAADF STEM imaging of sample G (grown at 850°C) is shown in Fig. 8(a). Uniform crystal structures are seen for both the MgSiN$_2$ and GaN layers. The interface between the MgSiN$_2$ film and GaN substrate is of extremely high quality, with an atomic layer interface. Fig. 8(b) illustrates a large-area STEM cross-section image of Sample G. Higher levels of non-uniformity in film thickness is evidenced for this sample, compared to Sample B. This result is consistent with the AFM results. The measured film thickness for Sample G ranges between 35 nm and 78 nm. Quantitative EDX measurements are plotted in Fig. 8(c). The Si and Mg compositions remain relatively uniform, within uncertainty, across the area, as also shown in the EDX color mapping of the elements in Fig. 8 (d)[(i)-(vi)]. The Mg and Si compositions are uniform. The Ga color mapping shows a clean interface. The oxygen composition in this film remains consistent across the sample, with no obvious elemental clustering.

Table 2 lists the STEM-EDX compositions of the films. A Mg:Si atomic ratio of 0.85 for Sample B, and 0.83 for Sample G were extracted from these measurements. Due to the sampling size and the area taken for SEM-EDX measurements (>1 µm x 1 µm), the SEM-EDX values are more accurate for the extracted Mg:Si ratio as compared to the relatively localized measurement obtained from STEM-EDX. The Mg:Si ratio obtained from STEM-EDX is within error of that extracted from SEM-EDX, showing consistency in extracted compositional ratios between both techniques. From these results, and the results described prior, potential for further improvements in crystalline quality and surface morphology is expected with further optimization of growth conditions.

## IV. Optical band gap extraction

Transmission measurements were performed on $MgSiN_2$ films grown on double-side polished, c-plane sapphire substrates. The sample growth conditions are listed in Table 3. Samples S and T were grown under the same conditions as Samples B and G, respectively. Figure 9(a) shows the ω-2θ scans for Samples S and T. Both films show (002) $MgSiN_2$ peaks. Figures 9(b) and 9(c) show AFM scans of Samples S and T, respectively. In both cases, the RMS values are smaller than for the corresponding films, Samples B and G, grown under the same conditions on GaN-on-sapphire substrates.

The transmittance measurements of these two $MgSiN_2$ films are shown in Fig. 10(a), with the Tauc plots shown in Fig. 10(b). The direct optical gaps of the $MgSiN_2$ films were determined to be 6.13±0.03 eV for Sample S and 6.27±0.03 eV for Sample T through extrapolation of a linear fit over the energy range 6.45 eV to 6.6 eV. The extracted direct gaps are within the expected range for the direct band gap of MgSiN$_2$ as predicted from DFT calculations (6.28 eV).[2] The indirect band gap is predicted to be at 5.84 eV. More detailed analysis to extract an indirect band gap will need to account for the presence of an Urbach tail, presumably dominated by cation exchange defects, and is the subject of future work.[18,9]

## IV. Conclusions

Single-crystal MgSiN$_2$ thin films were grown via MOCVD on GaN-on-sapphire substrates at 745 and 850 °C. The chamber pressures and the Mg:Si precursor molar flow rate ratios were tuned to optimize the quality of the MgSiN$_2$ films as measured by the MgSiN2 (002) XRD peak. The higher growth temperature resulted in a wider growth window, while the lower growth temperature yielded smoother films. No clear trends in the Mg:Si compositional ratio were observed with the variations in pressure. STEM results show the presence of a high-quality interface between the MgSiN$_2$ film and the underlying GaN template. In addition, single-crystal-oriented MgSiN$_2$ films were observed through HAADF STEM imaging. Extraction of the direct band gap from transmittance measurements of films grown on double-side-polished sapphire substrates yielded values of 6.13 eV and 6.27 eV, consistent with ab initio band structure calculations.

## Acknowledgements

This work was supported by Army Research Office (Award No. W911NF-24-2-0210).

## Author Declarations

### Conflict of Interest

The authors have no conflicts of interest to disclose.

**Data Availability**

The data that support the findings of this study are available from the corresponding author upon reasonable request.

† Thirupakuzi Vangipuram and Hu contributed equally to this work.

## Table Captions

**Table 1:** Growth conditions for samples grown on GaN-on-sapphire templates.

**Table 2:** EDX compositions for the MgSiN$_2$ film extracted from the STEM lamellas of Samples B and G. A Mg:Si composition ratio of 0.85 was obtained from EDX measurements for sample B. A Mg:Si composition ratio of 0.83 was obtained from EDX measurements for sample G.

**Table 3:** MOCVD growth conditions for Sample S and Sample T: MgSiN$_2$ films grown on c-plane, double-side polished sapphire substrates. Samples S and T were co-loaded with Samples B and G, respectively.

## Figure Captions

**Figure 1:** (a) XRD, (b) AFM, (c) SEM imaging and SEM-EDX Mg:Si composition of MgSiN$_2$ films grown on GaN-on-sapphire templates at 745°C with varied Mg:Si precursor molar flow rate ratios (samples A, B, C and D).

**Figure 2:** (a) XRD, (b) AFM and (c) SEM imaging and SEM-EDX Mg:Si composition of MgSiN$_2$ films grown on GaN-on-sapphire templates at 850°C with varied Mg:Si precursor molar flow rate ratios (samples E, F, G and H).

**Figure 3:** (a) Measured Mg:Si composition ratios as a function of varied Mg:Si precursor molar flow rate ratios, and (b) XRD film peak intensity to substrate peak intensity ratios as a function of varied Mg:Si precursor molar flow rate ratios.

**Figure 4:** (a) XRD, (b) AFM, and (c) SEM imaging and SEM-EDX Mg:Si composition of MgSiN$_2$ films grown on GaN-on-sapphire templates at 745°C at different growth pressures (samples I, J, K, L and M).

**Figure 5:** (a) XRD, (b) AFM and (c) SEM imaging and SEM-EDX Mg:Si composition of MgSiN$_2$ films grown on GaN-on-sapphire templates at 850°C at different growth pressures (samples N, O, P, Q and R).

**Figure 6:** (a) XRD film peak intensity to substrate peak intensity ratios as a function of varied growth pressure for samples grown at 745°C and 850°C and (b) AFM RMS surface roughnesses

of 5μm x 5μm scan areas as a function of varied growth pressure for samples grown at 745°C and 850°C.

**Figure 7:** (a) HAADF STEM cross-section of sample G showing the MgSiN$_2$ film grown on a GaN-on-sapphire template at 745°C. (b) TEM cross-section across a larger area for EDX mapping. (c) Quantitative elemental EDX elemental mapping across the MgSiN$_2$ film from top to bottom and (d) elemental EDX mapping for (i) Mg, (ii) Si, (iii) Ga, (iv) N, (v) O, and (vi) Au

**Figure 8:** (a) HAADF STEM cross-section of sample G showing the MgSiN$_2$ film grown on a GaN-on-sapphire template at 850°C. (b) TEM cross-section across a larger area for corresponding EDX mapping. (c) Quantitative elemental EDX elemental mapping across MgSiN$_2$ film from top to bottom and (d) Elemental EDX mapping for (i) Mg, (ii) Si, (iii) Ga, (iv) N, (v) O, and (vi) Au

**Figure 9:** (a) 2θ-ω XRD scans of samples S and T grown on double-side-polished sapphire substrates. (b) AFM scan of 5μm x 5 μm area for sample S and (c) AFM scan of 5 μm x 5μm area for sample T

**Figure 10:** (a) Transmittance measurements from 1.0 eV up to 6.7 eV for samples S and T and (b) Tauc plots for samples S and T.

**Table 1**

| Sample ID | Growth ID | Growth Temperature (°C) | Growth Pressure (Torr) | Mg:Si Precursor Molar Flow Rate Ratio | Measured SEM-EDX Mg:Si Composition Ratio | Measured AFM RMS Roughness (nm) |
|---|---|---|---|---|---|---|
| A | MgSiN#037 | 745 | 500 | 3.22 | 0.93±0.08 | 6.6±0.3 |
| B | MgSiN#033 | 745 | 500 | 3.30 | 0.92±0.08 | 5.0±0.3 |
| C | MgSiN#036 | 745 | 500 | 3.38 | 0.91±0.08 | 5.7±0.3 |
| D | MgSiN#040 | 745 | 500 | 3.46 | 1.06±0.09 | 11.8±0.6 |
| E | MgSiN#031 | 850 | 500 | 8.20 | 0.93±0.09 | 13.5±0.7 |
| F | MgSiN#038 | 850 | 500 | 8.50 | 0.99±0.09 | 14.6±0.7 |
| G | MgSiN#035 | 850 | 500 | 8.70 | 0.93±0.09 | 21.4±1.1 |
| H | MgSiN#039 | 850 | 500 | 9.00 | 1.19±0.11 | 17.3±0.9 |
| I | MgSiN#043 | 745 | 450 | 3.30 | 1.00±0.09 | 6.8±0.3 |
| J | MgSiN#047 | 745 | 475 | 3.30 | 1.10±0.09 | 6.1±0.3 |
| K | MgSiN#042 | 745 | 500 | 3.30 | 0.83±0.07 | 6.2±0.3 |
| L | MgSiN#049 | 745 | 525 | 3.30 | 0.92±0.08 | 3.7±0.2 |
| M | MgSiN#053 | 745 | 550 | 3.30 | 0.99±0.08 | 10.0±0.5 |
| N | MgSiN#045 | 850 | 450 | 8.70 | 1.11±0.10 | 9.2±0.5 |
| O | MgSiN#046 | 850 | 475 | 8.70 | 1.23±0.11 | 19.4±1.0 |
| P | MgSiN#044 | 850 | 500 | 8.70 | 0.91±0.08 | 23.9±1.2 |
| Q | MgSiN#048 | 850 | 525 | 8.70 | 1.11±0.10 | 21.5±1.1 |
| R | MgSiN#052 | 850 | 550 | 8.70 | 1.09±0.10 | 15.9±0.8 |

**Table 2**

| | Extracted EDX Composition from STEM | |
|---|---|---|
| | **Sample B** | **Sample G** |
| **Element** | **Atomic Percentage ± Error Percentage (%)** | **Atomic Percentage ± Error Percentage (%)** |
| Mg | 21 ± 4 | 23 ± 4 |
| Si | 25 ± 4 | 28 ± 4 |
| N | 50 ± 3 | 46 ± 3 |
| O | 2 ± 1 | 3 ± 1 |
| Ga | 0 ± 1 | 1 ± 1 |

**Table 3**

| Sample ID | Growth ID | Growth Temperature (°C) | Growth Pressure (Torr) | Mg:Si Precursor Molar Flow Rate Ratio | Extracted Optical Bandgap, $E_g$ (eV) |
|---|---|---|---|---|---|
| **S** | MgSiN#033 | 745 | 500 | 3.30 | 6.13 ± 0.03 |
| **T** | MgSiN#035 | 850 | 500 | 8.70 | 6.27 ± 0.03 |

**Figure 1**

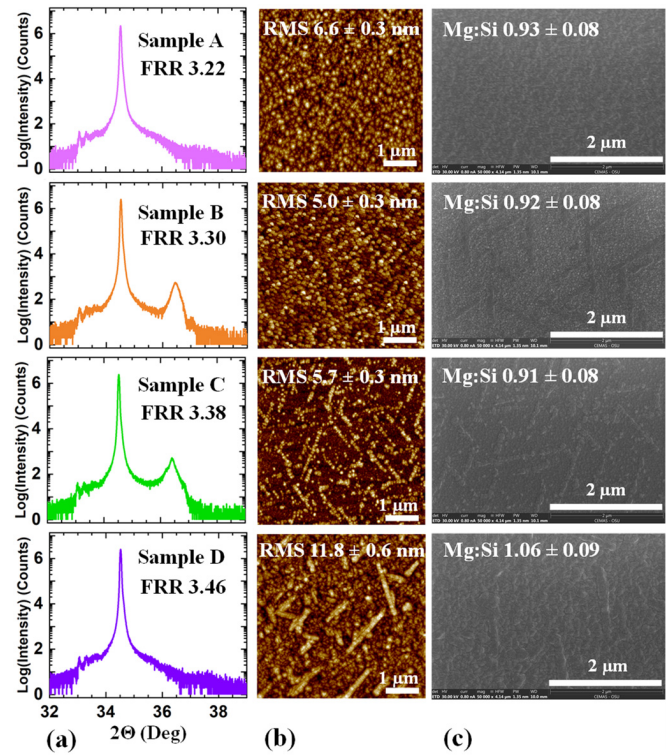

**Figure 2**

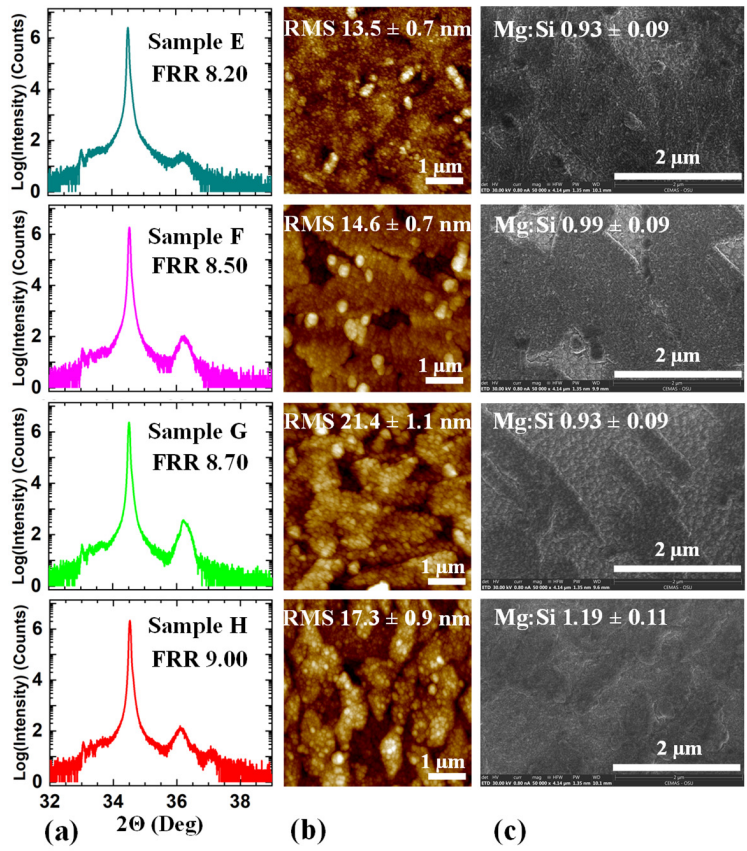

**Figure 3**

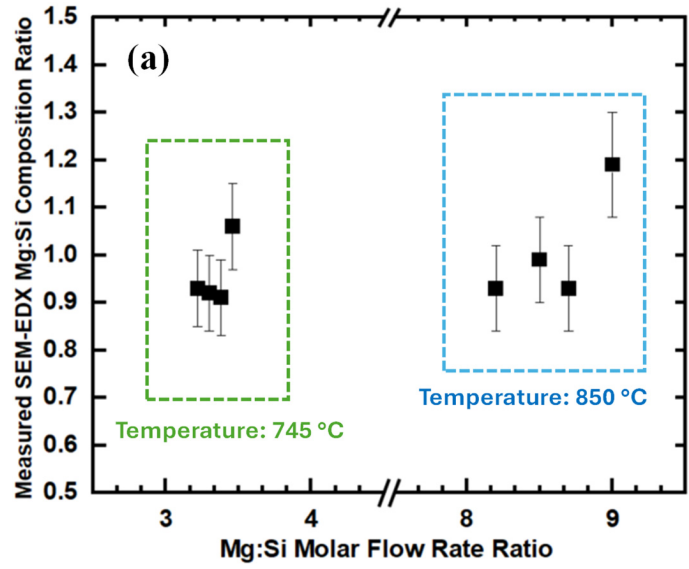

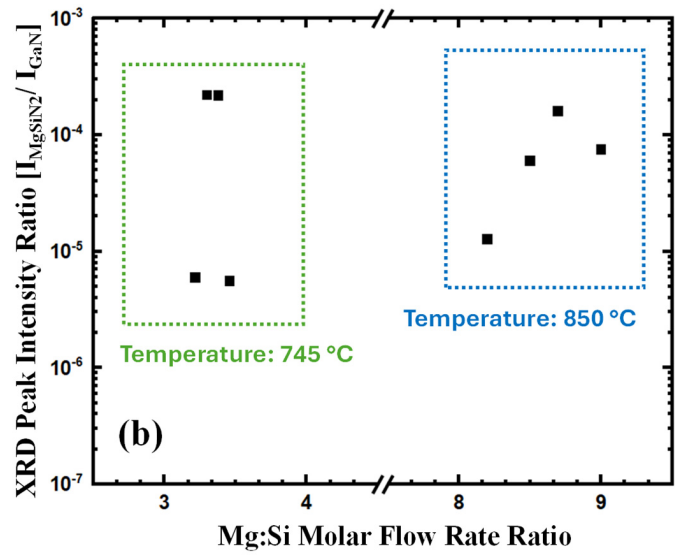

**Figure 4**

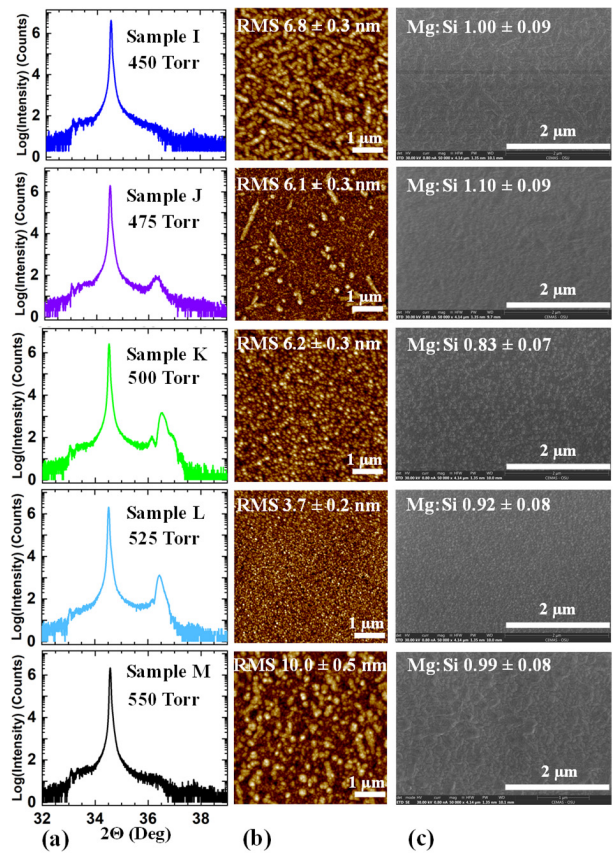

**Figure 5**

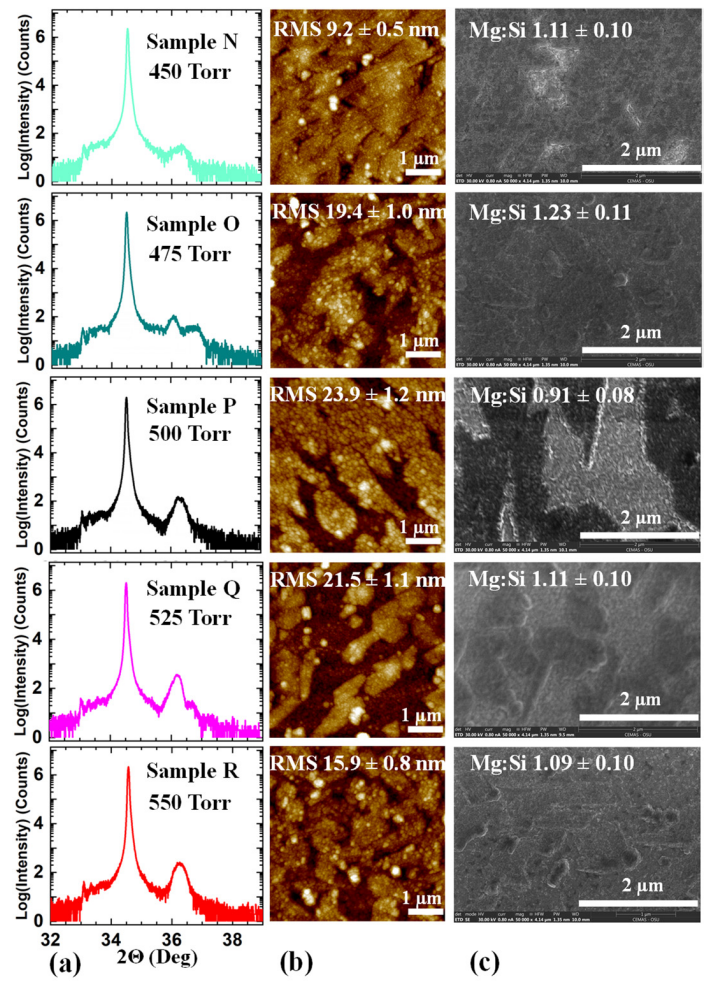

**Figure 6**

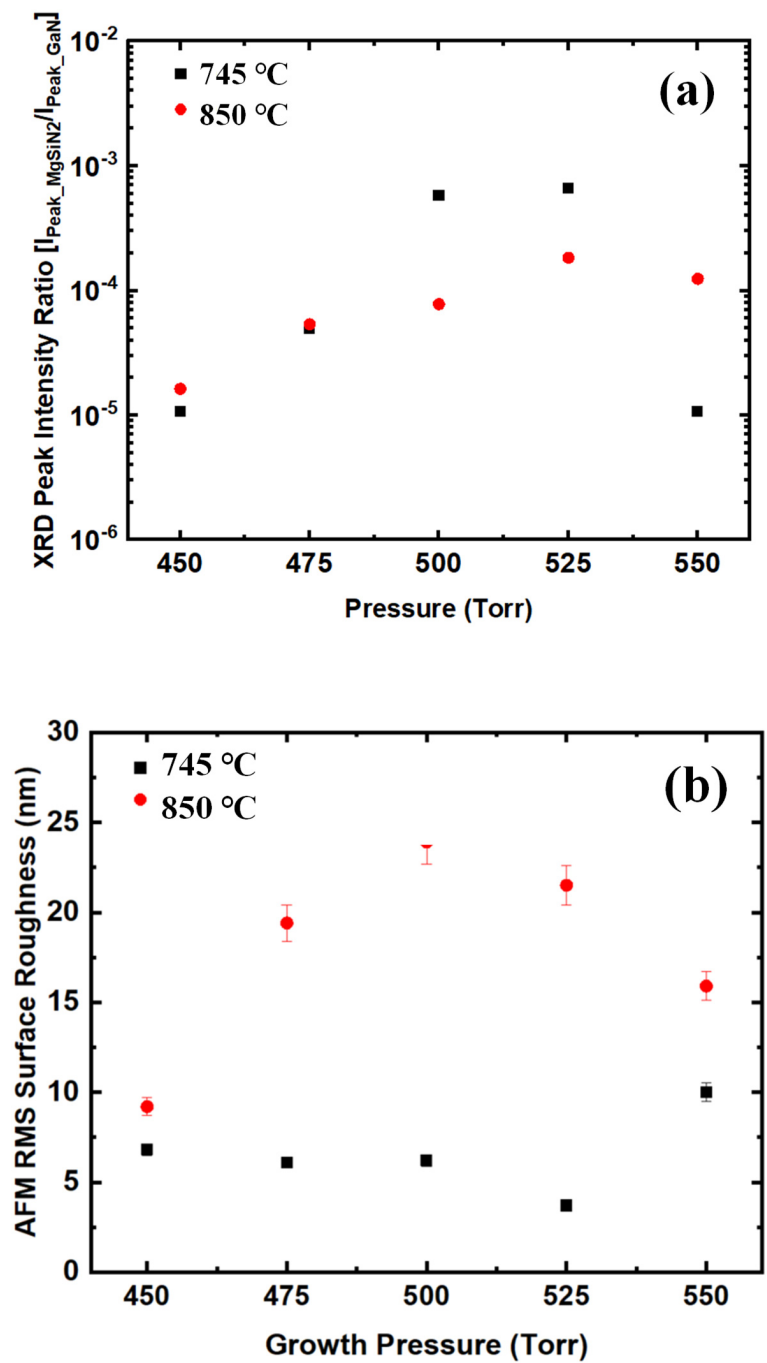

**Figure 7**

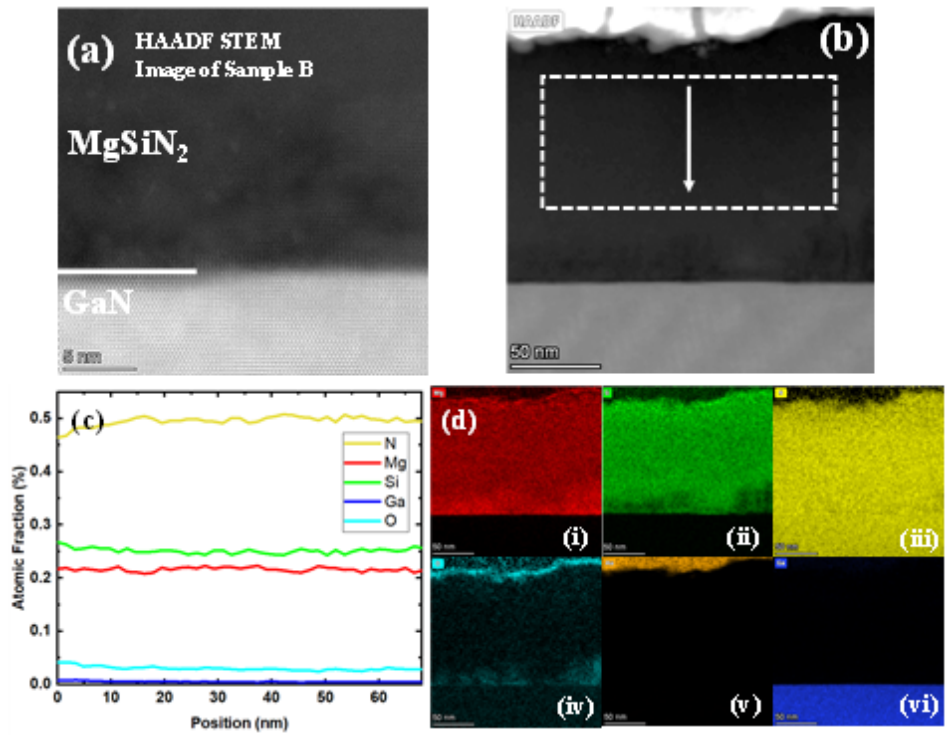

**Figure 8**

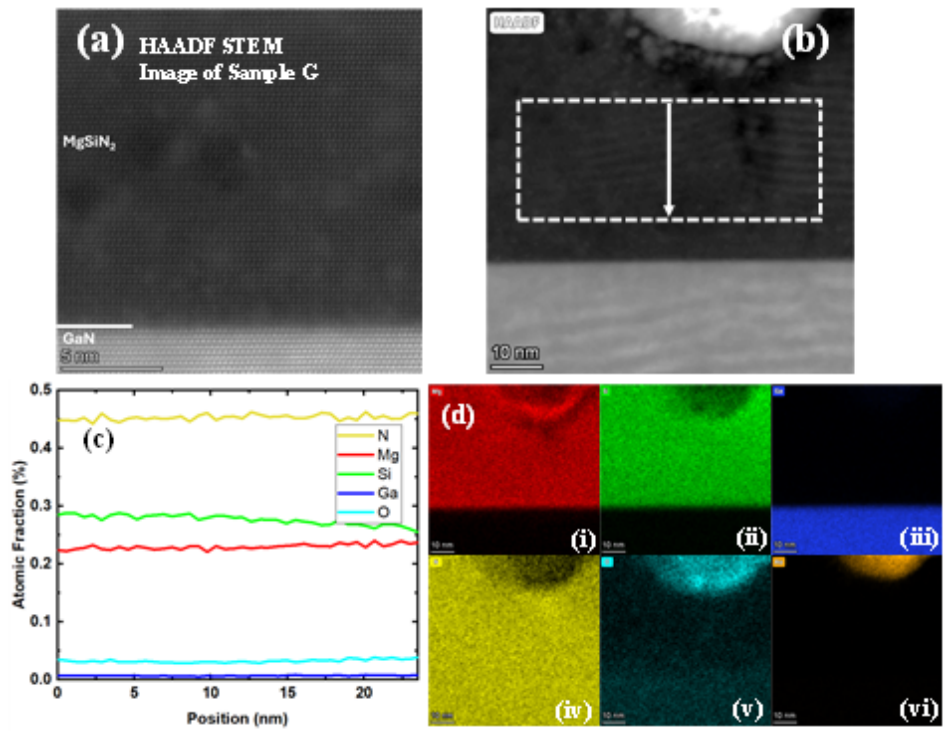

**Figure 9**

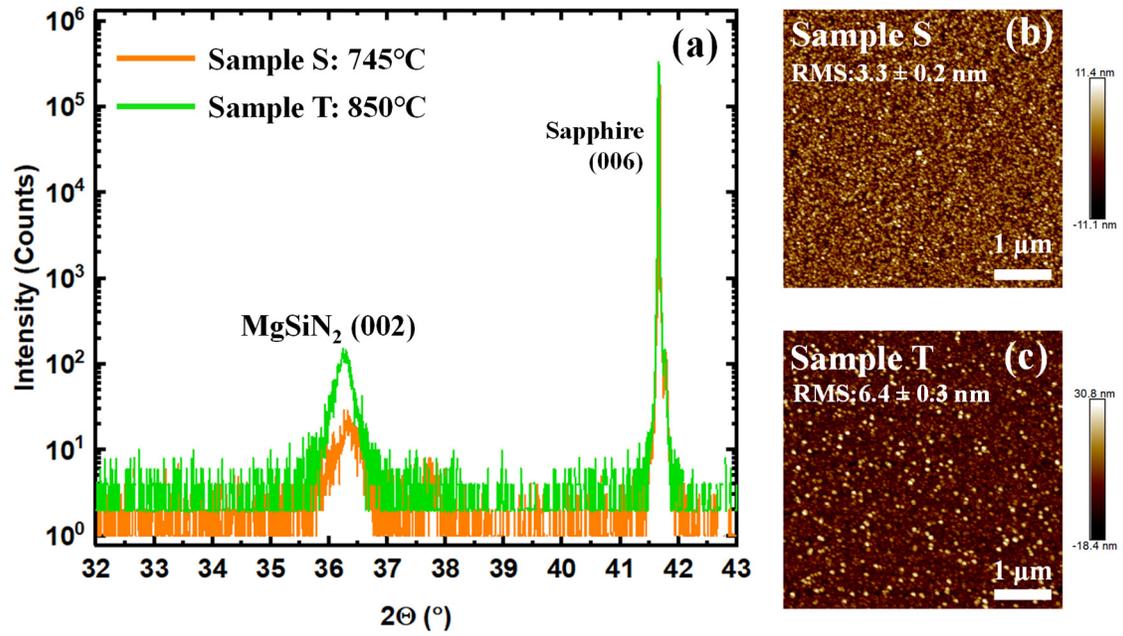

**Figure 10**

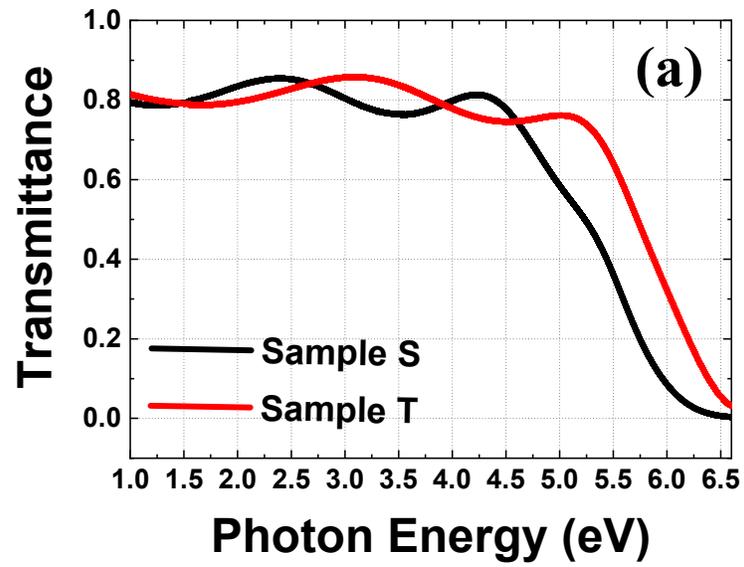

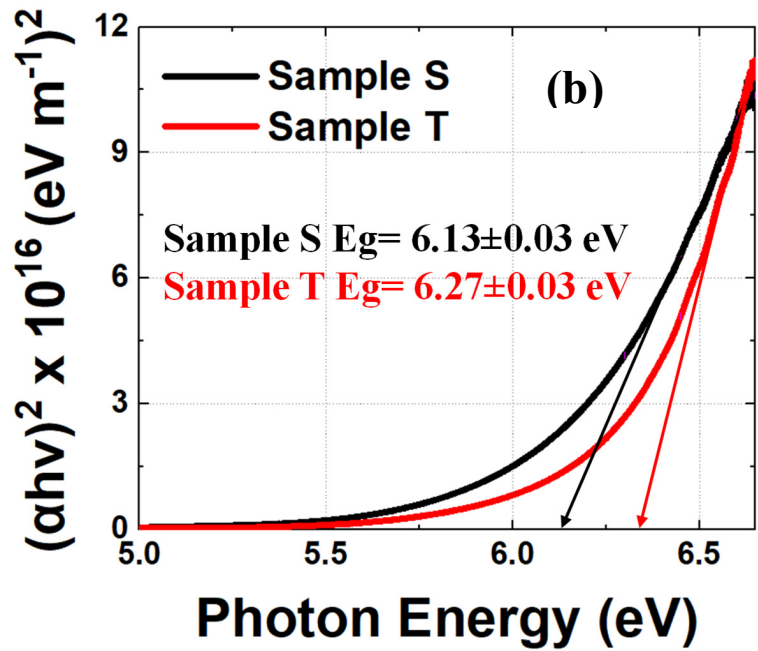